\documentclass[final,5p,times,twocolumn]{elsarticle}

 \usepackage{graphicx}
\usepackage{epsfig}

\usepackage{amssymb}

\newcommand{\br}{{\bf r}}

\newcommand{\be}{\begin{eqnarray}}
\newcommand{\ee}{\end{eqnarray}}
\newcommand{\bee}{\begin{eqnarray*}}
\newcommand{\eee}{\end{eqnarray*}}

\journal{Physics Letters B}

\begin{document}

\begin{frontmatter}



\title{Role of T=0 pairing in Gamow-Teller states in N=Z nuclei}


\author[label1]{C. L. Bai}
\address[label1]{Department of Physics, Sichuan University, Chengdu 610065, China}
\author[label2,label3]{H. Sagawa}
\ead{sagawa@u-aizu.ac.jp}
\address[label2]{Center for Mathematics and Physics, University of Aizu,
Aizu-Wakamatsu, 965-8580 Fukushima, Japan}
\address[label3]{RIKEN, Nishina Center,  Wako, 351-0198, Japan
}
\author[label3]{M. Sasano}
\author[label3]{T. Uesaka}
\author[label4]{K. Hagino}
\address[label4]{Department of Physics, Tohoku University, Sendai
980-8578, Japan}
\author[label5]{H. Q. Zhang}
\address[label5]{China Institute of Atomic Energy, Beijing 102413, China }
\author[label5]{X. Z. Zhang}
\author[label6]{F. R. Xu}
\address[label6]{School of Physics, Peking University, Beijing 100871, China}
\begin{abstract}
Gamow-Teller (GT) states in N=Z nuclei with the mass number A from 48 to 64 
are studied by using Hartree-Fock-Bogoliubov + quasi-particle
random phase approximation (HFB+QRPA) with Skyrme interactions.
The isoscalar spin-triplet (T=0,S=1) pairing  interaction is
taken into account in QRPA calculations.  It is found  
in the context of SU(4) symmetry in the spin-isospin space  that the GT strength
of lower energy excitations is largely  enhanced by
%
the
%
T=0 pairing interaction
which works cooperatively  with
%
the
%
T=1 pairing interaction in the ground state.
%
A two-peaked structure observed recently in $(p,n)$
 reaction on $^{56}$Ni 
can  be considered as  a manifestation of the role of 
T=0 pairing in the GT collective states.  
\end{abstract} 


\end{frontmatter}


The most prominent evidence  of pairing correlation in nuclei
%
is found
%
in the
odd-even staggering in binding energies and the gap in the excitation
 spectrum of
even-even nuclei 
in contrast to the compressed quasi-particle spectrum
in
odd-A nuclei \cite{BMP58,BM69,BB05}.
There are also dynamical effects of pairing
correlations seen in the moment of inertia associated with nuclear rotation and
large amplitude collective motion
\cite{BB05,Bertsch12,Matsuyanagi12}.
The Hartree-Fock (HF)+BCS method and  Hartree-Fock-Bogoliubov (HFB) method
have been commonly used to study the ground state properties of superfluid
nuclei
in a broad mass region \cite{Ben03,Stoi06,dug01,mar07}.  For the study of excited spectra, quasi-particle random phase approximation (QRPA)
has often been adopted as a basic method \cite{Engel,M01,Khan02,Paar03}.

The strong attraction between nucleons is the basic ingredient for
the pairing correlations. So far, the pairing interactions
of like-nucleons
with the isovector spin-singlet (T=1, S=0) channel has been mainly discussed.
In fact, the attraction between protons and neutrons is even stronger
in the isoscalar spin-triplet (T=0, S=1) channel \cite{Garrido}, which gives rise to the deuteron bound state.
However the  role of T=0 pairing  is limited in nuclei because of
large imbalance between neutron and proton numbers, and also the two-body
spin-orbit interaction which breaks the S=1 pair more effectively
than the S=0 pair \cite{Bertsch12,Poves98,Bertsch09}.
Nevertheless,  
the isoscalar pairing  causes extra binding energies 
in nuclei with N=Z 
 and has been considered  as one of 
the  origins of  the Wigner energy \cite{Satula97}.

Gamow-Teller (GT) states have been studied both experimentally and theoretically
intensively
in the last three decades.
Many interesting nuclear structure information has been revealed
by these studies, for example, the quenching of
sum rule strength \cite{Sakai} and the role of GT strength
in the astrophysical processes such as
neutrino-nucleus reactions \cite{Suzuki}.
 Because of recent development of modern
radioactive beam accelerator, it becomes feasible to observe GT states
in exotic nuclei near the proton and neutron drip lines.
   Recently, the GT transition strength was studied in a N=Z nucleus
 $^{56}$Ni which has an important impact on late stellar evolution through
electron capture and $\beta$ decay \cite{Sasano11}.
  Although the collective GT state is
mainly built of charge exchange particle-hole excitations, the low-lying
 strength responsible for $\beta$ decay involves neutron-proton $(np)$
particle-particle type excitations which could be sensitive to the T=0 pairing
interaction \cite{Engel,Moya}.  Intuitively,
  the  low-energy GT excitations  will be enhanced
by the following mechanism:
the T=1 pairing correlation provides partially occupied
proton and the neutron orbitals near the Fermi surface.  These orbitals
will accept the additional proton excitations  from the neutron orbits with the
same orbital $l$ quantum number.  Then,  these $(np)$ pair interacts through the T=0 pairing
interaction and enhances GT strength. 
These physical mechanisms may implement the super-allowed GT transition 
between the same multiplet members of  SU(4) symmetry in the spin-isospin space 
  which 
 never happened when  a
 neutron orbital is completely full and the corresponding proton one is empty.
 In this letter,
 we study GT states in several N=Z
nuclei focusing on the cooperative
role of T=1 and T=0 pairing interactions in the GT transitions. In Refs. \cite{Bai09,Bai10,Bai11},
  we applied successfully
 the self-consistent HF+RPA model to calculate spin and isospin
dependent excitations in several closed shell nuclei.
As an extension to open shell nuclei,
the HFB+QRPA method is used to calculate GT states
   with realistic  Skyrme
interactions.

The density dependent contact pairing interactions are
adopted for both T=1 and T=0 channels, 
\be
V_{T=1}(\br_1,\br_2)=V_0\frac{1-P_{\sigma}}{2}\Bigl(1-\frac{\rho(\br)}{\rho_o}\Bigr)\delta(\br_1-\br_2),\label{T1-pair}\\
V_{T=0}(\br_1,\br_2)=fV_0\frac{1+P_{\sigma}}{2}\Bigl(1-\frac{\rho(\br)}{\rho_o}\Bigr)\delta(\br_1-\br_2),\label{T0-pair}
\ee
where
$\br = (\br_1+\br_2)/2$ and
$\rho_0$ is the saturation density taken to be   $\rho_0$=0.16 fm$^{-3}$.
The strength of T=1 pairing is determined to reproduce
 the average odd-even mass staggering
 gap $\Delta^{(3)}$=12/A$^{1/3}$MeV in N=28 region 
 to be
$V_0$=$-550$MeV$\cdot$fm$^{3}$. 
We solve firstly  the HFB equations in coordinate space with a large radial mesh extending 
up to 20fm with a step of 0.05fm.
 In QRPA, we take 
into account all the states up to 180MeV excitation energy which have 
the occupation probabilities  more than $v^2=10^{-6}$.
The T=0 pairing is changed   from null  pairing to  strong pairing limit
 taking   the factor $f$ from $f$=0.0, 0.5, 1.0, 1.5 and 1.7.
The f=1.7 is the maximum value to avoid the spin-isospin instability of the
ground state of $^{56}$Cu.

\begin{figure}[t]
\begin{center}
\vspace{-0.8cm}
\includegraphics[clip,scale=0.42]{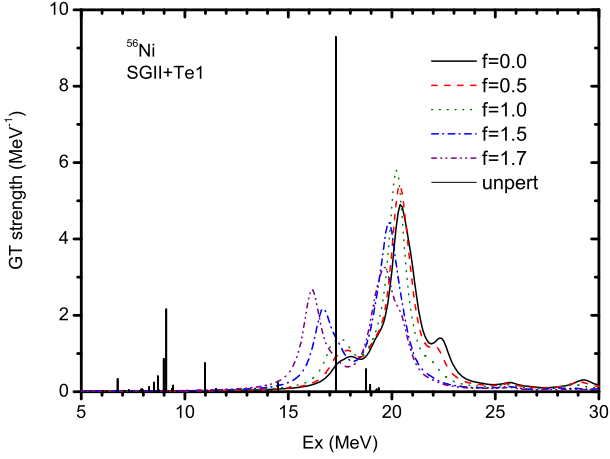}
\end{center}
\vspace{-0.5cm}
\caption{(Color online) Gamow-Teller strength in $^{56}$Ni obtained by HFB+QRPA
 calculations
with the Skyrme interaction  SGII+Te1.
The excitation energy is referred to the ground state of $^{56}$Ni.
  The T=0 pairing interaction is included in QRPA  changing the
coupling constant $f$ from  $f$=0.0, 0.5, 1.0, 1.5 and 1.7 in Eq.
  (\ref{T0-pair}).
The solid lines show the unperturbed strength without RPA correlations.
The QRPA strength is smoothed out by a Lorentzian function
with a width $\Gamma=1$ MeV} \label{fig1}
\vspace{-0.3cm}
\end{figure}

The GT transition involves both the spin and isospin operators as 
\be \label{GT}
\hat{O}(GT)=\sum_i {\bf \sigma}(i) t_{\pm}(i)  
\ee
for spin-flip and isospin-flip excitations.
 The calculated GT states in $^{56}$Ni  for $t_{-}$ channel are shown in Fig.\ref{fig1}
where the interaction SGII+Te1 \cite{Bai10,Bai11} is used
 for  HF mean field and particle-hole interactions.  The unperturbed strengths shown by
solid straight lines have  two groups at E$_x$=9 MeV and 17 MeV.  The QRPA results without
T=0 pairing  shown by the black curve have   the
 main peak around 20.4 MeV together
with a small peak at 18.0 MeV.
When the T=0 pairing strength is introduced,
the global GT strength is shifted to lower energy. At the same time,
the GT strength of lower peak is increased and absorbs the
 major strength when the T=0 pairing
is stronger than T=1 ($f\geq 1$).

\begin{figure}[t]
\begin{center}
\vspace{-1.cm}
\includegraphics[clip,scale=0.55]{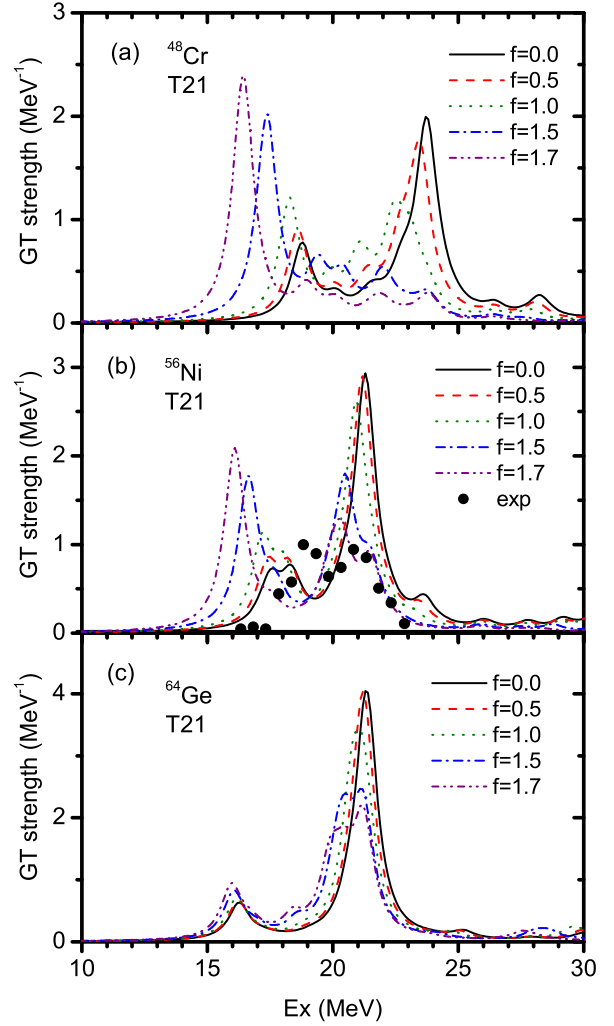}
\vspace{-0.8cm}
\caption{(Color online) Gamow-Teller strength in $^{48}$Cr, $^{56}$Ni and
$^{64}$Ge by HFB+QRPA
with
the Skyrme interaction T21.
The quenching factor 0.74 is adopted
  for the GT transition operator $\hat{O}($GT).  
 The T=0 pairing interaction is included in QRPA calculations
by changing the
coupling constant  from  $f$=0.0, 0.5, 1.0, 1.5 and 1.7 in Eq. (\ref{T0-pair})
The experimental data of $^{56}$Ni are taken from Ref. \cite{Sasano11}.
} \label{fig:T21}
\end{center}
\vspace{-0.5cm}
\end{figure}

The HFB+QRPA results for $t_{-}$ channel 
are shown in Fig. \ref{fig:T21} for $^{48}$Cr,
$^{56}$Ni and $^{64}$Ge. The interaction T21 \cite{Lesinski07}  is used
for  HF mean field and particle-hole interactions.  
When the T=0 pairing strength is introduced, 
the global GT strength is shifted to lower energy.   
At the same time,  
 the increase of lower energy peak is always seen in the three nuclei
as the T=0 pairing becomes stronger.
As a systematic trend, the lighter the mass is
the larger the lower  energy GT peak is.
In the two nuclei  $^{56}$Ni and $^{64}$Ge ,
the double peak structure
can be seen even in the strong T=0 pairing limit at $f=1.7$.  On the contrary,
  in $^{48}$Cr,
  the higher energy peak
 is almost disappeared in the case of  $f=1.7$. We found this trend also
  in $^{44}$Ti in which
 the higher energy peak is very small and is not seen 
 in the case of $f\geq 1.5$.
 
The experimental data of $^{56}$Ni show two peaks at 18.7 and 20.8 MeV
 \cite{Sasano11}, as is shown in the middle panel of Fig. 2.
The calculated results for $^{56}$Ni with T=0 pairing of 
 $f=1.0$ and $f=1.5$ give almost the
identical
peak energy to the observed one for 
 the higher excitation  peak.
   On the other hand,  
   the calculated lower  peak is 2 MeV lower
than the experimental one  in energy.

\begin{table}[t]
\caption{Amplitudes of main $(np)$ particle-hole  and
 particle-particle type configurations  of GT states in $^{56}$Ni.
The QRPA calculations are performed without and with the T=0 pairing interaction
 in the cases of $f=0$ and $f=1.5$, respectively. The Skyrme interaction
T21 is used for HF and p-h matrix calculations.
The abbreviations $B$ and $C$ correspond to the GT
 reduced matrix element
and
the normalization factor defined in Eqs. (\ref{B}) and (\ref{RPA-amp}), respectively.
 The excitation energy Ex is given in unit of 
MeV \label{Tab1}.}
\begin{center}
\begin{tabular}{c|c|cc|cc}
\hline
 \multicolumn{2}{c}{$^{56}$Ni} & \multicolumn{2}{c}{$f=0$} &  & \\ \hline
 E$_x$  & B(GT) &$\nu(v_{\nu}^2)$ & $\pi(v_{\pi}^2)$ &  B & C     \\ \hline
 17.5 & 1.41 & 2p$_{3/2}(0.20)$ & 2p$_{3/2}(0.21)$ & 0.358  & 0.127 \\
  & & 1f$_{7/2}(0.69)$ &  1f$_{5/2}(0.09)$ &   $-$0.229   & 0.006   \\
  & &  1f$_{7/2}(0.69)$ &  1f$_{7/2}(0.67)$ &  1.341 & 0.802  \\ \hline
 18.3& 1.49 & 2p$_{1/2}(0.10)$ &  2p$_{3/2}(0.21)$ & 0.260 & 0.153 \\
    & & 2p$_{3/2}(0.20)$ &  2p$_{1/2}(0.11)$ & 0.846 & 0.740 \\\hline
 21.3 & 7.88&  1f$_{7/2}(0.69)$ &  1f$_{5/2}(0.09)$ & 2.48  & 0.742\\ \hline
  \multicolumn{2}{r}{S$_-$(GT)=18.28}& \multicolumn{4}{c}{}\\ \hline
\multicolumn{2}{c}{$^{56}$Ni} & \multicolumn{2}{c}{$f=1.5$} &  & \\ \hline
E$_x$ & B(GT) &$\nu(v_{\nu}^2)$ & $\pi(v_{\pi}^2)$ &  B & C     \\ \hline
 16.6 & 4.82 & 2p$_{3/2}(0.20)$ & 2p$_{1/2}(0.11)$ & $-$0.203 & 0.049 \\
 & & 2p$_{3/2}(0.20)$ & 2p$_{3/2}(0.21)$  &  $-$0.682  & 0.491  \\
  & & 1f$_{7/2}(0.69)$ &  1f$_{5/2}(0.09)$ &   $-$0.237 & 0.007  \\
  & &  1f$_{7/2}(0.69)$ &  1f$_{7/2}(0.67)$ &  $-$0.790 & 0.339 \\ \hline
 20.5& 4.10
&  1f$_{7/2}(0.69)$ &  1f$_{5/2}(0.09)$ &  $-$1.900 & 0.437 \\\hline
 \multicolumn{2}{r}{S$_-$(GT)=14.75}& \multicolumn{4}{c}{}\\
\hline
\end{tabular}
\end{center}
\vspace{-0.5cm}
\end{table}%

 Main  configurations of  dominant GT states in  $^{56}$Ni
are tabulated  in Table \ref{Tab1}  without and with the
T=0 pairing interaction
($f=0.0$ and $f=1.5$), respectively.  In Table \ref{Tab1}, the value  
 B corresponds to the GT
 reduced matrix element
\be  \label{B}
\mbox{B}=(Xu_{\pi}v_{\nu}-Yu_{\nu} v_{\pi})\langle\pi||\hat{O}(GT)||\nu\rangle
\ee
 with the operator Eq. (\ref{GT}),   and
the value
 \be  \label{RPA-amp}
\mbox{C}=X^2-Y^2, 
\ee
is the  normalization factor where $X$ and $Y$
are QRPA amplitudes.  The dominant configurations are only listed  in Table 1.  
Without the T=0 pairing,
 the main configuration
of higher excitation is
a particle-hole type (1f$_{7/2}^{\nu}$,1f$_{5/2}^{\pi}$) excitation referring to the configuration
of fully occupied 1f$_{7/2}$ proton and neutron  orbits, while that of
lower energy has   particle-particle type
(1f$_{7/2}^{\nu}$,1$f_{7/2}^{\pi}$), (2p$_{3/2}^{\nu}$,2p$_{3/2}^{\pi}$) and
 (2p$_{3/2}^{\nu}$,2p$_{1/2}^{\pi}$)  excitations.  These particle-particle type excitations   become
possible only when   T=1 pairing interaction acts on  the ground state.
  The occupation probabilities of neutron
 single particle states are obtained by integrating the lower components of HFB wave functions as 
$v^2$=0.69, 0.20 , 0.10 and 0.08 for 1f$_{7/2}$, 2p$_{3/2}$ ,
2p$_{1/2}$ and 1f$_{5/2}$ orbits, respectively.  The proton states
have almost the same occupation probabilities.
 With a strong T=0 pairing ($f$=1.5),
 the main configurations of the low energy  peak
 become  a coherent
superposition of 
two quasi-particle (QP)
excitations  (2p$_{3/2}^{\nu}$,2p$_{1/2}^{\pi}$),
 (2p$_{3/2}^{\nu}$,2p$_{3/2}^{\pi}$), (1f$_{7/2}^{\nu}$,1f$_{5/2}^{\pi}$) and (1f$_{7/2}^{\nu}$,1$f_{7/2}^{\pi}$)
as shown in Table \ref{Tab1}.
These  coherent phases are inherent to the T=0 pairing interaction and
increases the GT strength.  On the other hand,  the higher energy peak is  still  dominated by the (1f$_{7/2}^{\nu}$,1f$_{5/2}^{\pi}$) configuration, but
 decreases its strength significantly.

The GT sum rule values S$_- ($GT) are
 given  in  Table \ref{Tab1} 
without and with the T=0 pairing ($f=0.0$ and $f=1.5$).
The  S$_- ($GT) value is decreased when the T=0 pairing is included
\cite{Moya}.
This is due to the fact that
 the high excited GT strength is more deceased than the increase
of the lower excited GT strength since the main configuration of GT strength
(1f$_{7/2}^{\nu}$,1f$_{5/2}^{\pi}$) is hindered.  The decrease of GT
strength is always the case with T=0 pairing
also in the other N=Z nuclei studied in the present paper.
 The decrease of the  S$_- ($GT)  value is about 20\%  and going  up
   to 30\% in the cases of $f=1.5$ and  $f=1.7$, respectively.
 We  have checked the increase of GT strength in the lower peak in $^{56}$Ni
by using other Skyrme interactions SGII+Te1\cite{Bai10,Bai11}, T43 and T32  \cite{Lesinski07}.
We found  that  these features of GT states are
  robust with some minor differences in the excitation energy
of two main peaks.

The enhancement of lower energy GT strength by the
   (T=0, S=1) pairing can be induced by the transition
between four QP states  such as
\bee  \label{4QP}
\hspace{-0.5cm}
\langle\{(j^2 )^{T=1}_{S=0}(p_{3/2}^{\nu}p_{3/2}^{\pi})^{T=0}_{S=1}\}^{T_f=1}_{S_f=1}|| \hat{O(GT)}||\{(j^2)^{T=1}_{S=0}(p_{3/2}^{\nu}p_{3/2}^{\nu})^{T=1}_{S=0}\}
^{T_i=0}_{S_i=0}\rangle  
\nonumber \\
\hspace{-2cm} =\langle (p_{3/2}^{\nu}p_{3/2}^{\pi})^{T=0}_{S=1}||\hat{O(GT)}||(p_{3/2}^{\nu}p_{3/2}^{\nu})
^{T=1}_{S=0}\rangle  \,\,\,\,\,\,\, (6)
\eee
where the two quasi-particle configuration $(j^2)^{T=1,S=0}$ 
(for examples,  $j=1f_{7/2}^{\nu}$ or $j=2p_{3/2}^{\nu}) $
 acts as a spectator. The double bar in Eq.
  (\ref{4QP})
is the symbol of doubly reduced matrix element both in spin and isospin space.
Other  QP configurations $(1f_{7/2}^{\nu},1f_{7/2}^{\pi})^{T=0,S=1}$,   $(1f_{7/2}^{\nu},1f_{5/2}^{\pi})^{T=0,S=1}$ and
$(2p_{3/2}^{\nu}2p_{1/2}^{\pi})^{T=0,S=1}$ are  also
  possible configurations in Eq. (\ref{4QP})
equivalent to  $(2p_{3/2}^{\nu}2p_{3/2}^{\pi})^{T=0,S=1}$.
It should be noticed that the T=1 pairing works between the two particle
 configurations
$|(lj)^2:(L=S=0)J=0,T=1\rangle$ with $j=7/2,5/2,3/2$ and $1/2$ in $pf$
  shell model space.  On the other hand, the T=0 pairing acts on the configurations
$|(lj,l'j'):(L=0,S=1)J=1,T=0\rangle$
with not only ($l=l'$, $j=j'$), but also with  ($l=l'$,
 $j=j'\pm 1$).  This is the  reason why
the T=0 pairing makes  strong couplings among (T=0,S=1) pairs with
$(2p_{3/2}^{\nu}2p_{1/2}^{\pi})$ and  $(1f_{7/2}^{\nu}1f_{5/2}^{\pi})$ 
 as well as $(2p_{3/2}^{\nu}2p_{3/2}^{\pi})$ 
  and $(1f_{7/2}^{\nu}1f_{7/2}^{\pi})$, enhancing  the lower energy GT  peak
   in the case of $f=1.5$ in
 Table \ref{Tab1}.  

Main configurations of GT states in $^{48}$Cr and $^{64}$Ge are listed in Table \ref{Tab2} calculated with a strong T=0 pairing ($f=1.5$).  Because of partially occupied
 2p$_{3/2}$,
 1f$_{7/2}$ and 1d$_{3/2}$ orbits in $^{48}$Cr,
  the configurations (2p$_{3/2}^{\nu}$2p$_{3/2}^{\pi}$), (1f$_{7/2}^{\nu}$1f$_{7/2}^{\pi}$) and
  (1d$_{3/2}^{\nu}$1d$_{3/2}^{\pi}$)  give large
contributions to the lower energy GT
 strength at E$_x$=17.4 MeV having  coherent phases.
 Although  the high energy state at  E$_x$=19.4 MeV obtains  large
 contributions from (2p$_{3/2}^{\nu}$2p$_{3/2}^{\pi}$) and 
 (1f$_{7/2}^{\nu}$1f$_{5/2}^{\pi}$), it is  cancelled  by
 other configurations   (1d$_{3/2}^{\nu}$1d$_{3/2}^{\pi}$) and (1f$_{7/2}^{\nu}$1f$_{7/2}^{\pi}$).  This cancellation
decreases substantially the B(GT) value compared with that of the state at
E$_x$=17.4 MeV.  The situation is rather different in $^{64}$Ge since
neutron and proton 1f$_{7/2}$  orbits are almost fully occupied.
 Thus the (1f$_{7/2}^{\nu}$1f$_{7/2}^{\pi}$)  configuration does not appear as
a dominant configuration in the strong B(GT) states in Table II. 
 On the other hand,   the 2p$_{1/2}$,
  2p$_{3/2}$ orbits contribute significantly to the lower energy state at
  E$_x$=16.1 MeV in $^{64}$Ge.
The B(GT) value of this state in $^{64}$Ge is only a half of the state at E$_x$=17.4 MeV
 in $^{48}$Cr because of negligible contribution from
the (1f$_{7/2}^{\nu}$1f$_{7/2}^{\pi}$) configuration
even though the T=0 pairing is strong.
The high energy state at  E$_x$=21.2 MeV in $^{64}$Ge is dominated by the
 (1f$_{7/2}^{\nu}$1f$_{5/2}^{\pi}$)  configuration,
the same as the high energy state  in  $^{56}$Ni.  The sum rule values S$_-$(GT) of $^{48}$Cr and $^{64}$Ge are
also given in Table \ref{Tab2} in the case of strong T=0 pairing.
In general,  the
sum rule value is larger for the heavier N=Z nuclei since available
configuration space is larger.  The increase of
S$_-$(GT) is 25\% and 47\% in  $^{56}$Ni and $^{64}$Ge, respectively,
in comparison with that of  $^{48}$Cr.

\begin{table}[htp]
\caption{Same as Table  \ref{Tab1},  but
for $^{48}$Cr and $^{64}$Ge with 
the  T=0 pairing interaction
 $f=1.5$.
 \label{Tab2}}
\begin{center}
\renewcommand{\arraystretch}{1.5}
\begin{tabular}{c|c|cc|cc}
\hline
\multicolumn{2}{c}{$^{48}$Cr} & \multicolumn{2}{c}{$f=1.5$} &  & \\ \hline
E$_x$  & B(GT) &$\nu$(v$_{\nu}^2)$ & $\pi$(v$_{\pi}^2)$ &  B & C     \\ \hline
 17.4& 5.68 & 2p$_{3/2}(0.12)$ & 2p$_{3/2}(0.12)$ & 0.186  & 0.062 \\
 & & 1d$_{3/2}(0.85)$ & 1d$_{3/2}(0.84)$  &  0.251 & 0.204 \\
 & &  1f$_{7/2}(0.33)$ &  1f$_{5/2}(0.07)$ &  0.268  & 0.021 \\
 & &  1f$_{7/2}(0.33)$ &  1f$_{7/2}(0.32)$ &  1.04 & 0.558 \\\hline
 19.4& 1.19 & 2p$_{3/2}(0.12)$ & 2p$_{1/2}(0.07)$ & 0.215 & 0.081 \\
 & &  2p$_{3/2}(0.12)$ & 2p$_{3/2}(0.12)$ & 0.559  & 0.461 \\
 & & 1d$_{3/2}(0.85)$ & 1d$_{3/2}(0.84)$  &  $-$0.217 & 0.142 \\
 & &  1f$_{7/2}(0.33)$ &  1f$_{5/2}(0.07)$ &  0.383 & 0.038 \\
 & &  1f$_{7/2}(0.33)$ &  1f$_{7/2}(0.32)$ &  $-$0.384 & 0.054 \\\hline
 \multicolumn{2}{r}{S$_-$(GT)=11.77}& \multicolumn{4}{c}{}\\ \hline
\multicolumn{2}{c}{$^{64}$Ge} & \multicolumn{2}{c}{$f=1.5$} &  & \\ \hline
E$_x$  & B(GT) &$\nu$(v$_{\nu}^2)$ & $\pi$(v$_{\pi}^2)$ &  B & C     \\ \hline
 16.1& 2.15 &
 2p$_{3/2}(0.48)$ & 2p$_{1/2}(0.21)$ & $-$1.316 & 0.889\\
\hline
 21.2 & 5.21 & 1f$_{5/2}(0.14)$ &  1f$_{7/2}(0.90)$ & $-$0.187& 0.353 \\
& &  1f$_{7/2}(0.92)$ &  1f$_{5/2}(0.15)$ & 2.392& 0.561 \\\hline
 \multicolumn{2}{r}{S$_-$(GT)=17.26}& \multicolumn{4}{c}{}\\
\hline
\end{tabular}
\end{center}
\vspace{-0.5cm}
\end{table}%

\begin{figure}[t]
\vspace{-0.5cm}
\includegraphics[clip,scale=0.42]{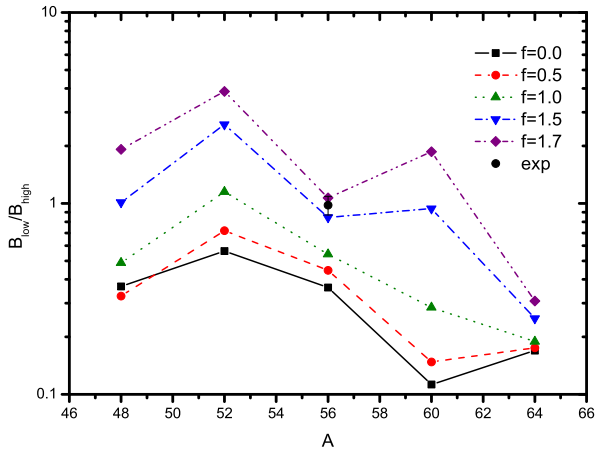}
\caption{(Color online) Ratio between
Gamow-Teller peaks in the lower energy and the higher energy regions
as a function of the T=0 strength parameter, $f$,  calculated 
 with T21 interaction.
The lower energy and higher energy region is divided  at E$_{x}$=20MeV.
The experimental ratio in   $^{56}$Ni is obtained from observed  B(GT)
strengths in the lower and higher energy regions  in Ref. \cite{Sasano11}.
} \label{B-ratio}
\vspace{-0.3cm}
\end{figure}

The peak ratio between  GT strength of the low energy region (below E$_{x}$=20 MeV) and the
high energy region (above E$_{x}$=20 MeV)
is plotted in Fig. \ref{B-ratio} as a function of mass number $A$.
In the cases of  weak T=0 pairing ($f\leq 1.0$), the high energy peak
 B$_{\rm high}$(GT) is larger than the low energy one  B$_{\rm low}$(GT).
With  stronger T=0 pairing ($f>1.0$), the
B$_{\rm low}$(GT) value becomes larger and dominates the GT strength distribution, especially
in nuclei $A\leq56$.  The low energy peak may appear  as a single giant resonance in the strong
T=0 pairing limit.  The empirical ratio  of B$_{\rm low}$(GT) to B$_{\rm high}$(GT) in $^{56}$Ni
is  obtained from observed B(GT)
strength distributions in Ref. \cite{Sasano11}.  The data point is consistent with the calculated
 ratio  with the T=0 pairing at $f\sim 1.5$.
 
The shell model calculations were performed for  $^{56}$Ni by using effective 
interactions KB3G and GXPF1A in ref.~\cite{Sasano11,SH12} and a  particle-vibration coupling 
model is applied to calculate the GT strength in $^{56}$Cu in ref. \cite{Niu12}.  
The shell model interaction GXPF1A gives a two-peak structure
 for GT strength distributions.  The competition between T=0 pairing and the 
 spin-orbit interaction may be a key issue of these results~\cite{SH12}.  
It is also shown in ref. \cite{Niu12} that   the effect of particle-vibration coupling 
gives a sizable effect on  GT strength distribution in  $^{56}$Cu based on the  HF+RPA model. 
It might be an interesting and 
challenging task to take into account the effect beyond HFB+QRPA such as the 
 particle-vibration coupling effect 
in the microscopic model in future.



The strength of the T=0 pairing can be determined by using the $p$-$n$
scattering length \cite{LW74}
in the same way as for the T=1 pairing \cite{BE,EBH97}.
The extracted value is slightly larger than the T=1 pairing strength and,
makes a bound deuteron state with a binding energy of $1.41$MeV,
which is somewhat smaller than the experimental
value  $2.2$MeV.
The effective density dependent pairing forces of zero range are
obtained in T=0 and 1 channels in symmetric nuclear matter from
Paris forces in Ref. \cite{Garrido}. The obtained T=0 interaction
 is much stronger than the T=1 interaction.
In Ref. \cite{Nicole}, a phenomenological T=0 pairing strength was determined
  to be 1.36 times stronger than the corresponding T=1 pairing by fitting
the ground state spins
of several N=Z  nuclei in the p shell region. The large value of the
T=0 pairing can be considered
to mimic the effect of tensor correlations which contributes to the
binding energy of deuteron
significantly.   At present, it is still an open question how much is
 the phenomenological  T=0 pairing interaction,  which can be used for
HFB and QRPA calculations in medium-heavy and also in heavy nuclei.
As we have pointed out in this letter, the empirical data of GT strength
in N=Z  nuclei  will provide important information to disentangle the T=0 pairing  in
medium-heavy and heavy mass nuclei.

In summary, we study GT states in several N=Z nuclei in medium mass region by using HFB+QRPA model
taking into account T=0 pairing interaction.  We found
in the context of SU(4) supermultiplets that the low excitation energy
GT strength is largely enhanced by T=0 pairing interaction which is
working cooperatively  with the T=1 pairing
correlations in the ground state.  
Especially in the lighter nuclei N=Z=24$\sim$28, the
B$_{\rm low}$(GT)
becomes comparable or even larger than the B$_{\rm high}$(GT) values in the strong T=0 pairing
$(f\geq1.0)$.  This enhancement of the lower energy GT strength is attributed to the coherent excitation  of several $pf$ shell configurations
 and   can be considered as an implementation 
of the super-allowed GT transition in the SU(4) supermultiplets 
in the  spin-isospin space.
    It is found that the available data in $^{56}$Ni is consistent with
the calculated ratio B$_{\rm low}$(GT)/B$_{\rm high}$(GT) with
 the T=0 pairing strength $f\sim 1.5$.
Further additional information of empirical
B(GT) is desperately desired to confirm the role of T=0 pairing in the GT states in N=Z nuclei.

We are grateful to Toshio Suzuki, M. Honma, M. Ichimura, 
  K. Yako and T. Wakasa for valuable discussions.  This work is supported by the National Natural Science Foundation of
China under Grant Nos. 10875172, 10275092,10675169,and 11105094,
and Financial Supports from Sichuan University (Project Nos.
2010SCU11086 and 2012SCU04A11). This work was also  supported by the
Japanese Ministry of Education, Culture, Sports, Science and
Technology by Grant-in-Aid for Scientific Research under the program
numbers  (C) 22540262.





\bibliographystyle{elsarticle-num}

\begin{thebibliography}{00}

\bibitem{BMP58} A. Bohr, B. R. Mottelson, and D. Pines, Phys. Rev.
{\bf 110}, 936 (1958).
\bibitem{BM69}  A. Bohr and  B. R. Mottelson,  Nuclear Structure(Benjamin, New York, 1969) Vol. I.
\bibitem{BB05} D. M. Brink and R. Broglia, \textit{``Nuclear superfluidity,
pairing in Finite Systems"}, Cambridge monographs on particle physics, nuclear
physics and cosmology, vol 24 (2005).
\bibitem{Bertsch12} G. F. Bertsch, \textit{`` 50 Years of Nuclear BCS''},
 (World Scientific, edited by R. Broglia and V. Zelevinsky), 
  arXiv:1203.5529v1 (March,2012).

\bibitem{Matsuyanagi12}
K. Matsuyanagi, N. Hinohara, and K. Sato,
arXiv:1205.0078v1 (May,2012).

\bibitem{Ben03}  M. Bender, P.-H.Heenen and P.-G. Reinhard,  Rev. Mod. Phys.
  {\bf 75}, 121 (2003).
\bibitem{Stoi06} M. V. Stoitsov, J. Dobaczewski, W. Nazarewicz and P. Borycki,
  Int. J. Mass Spectrum, {\bf 251}, 243 (2006).
\bibitem{dug01} T. Duguet, P. Bonche, P.-H. Heenen, and J. Meyer,
  Phys. Rev. C {\bf 65}, 014311 (2001).
\bibitem{mar07}  J. Margueron, H. Sagawa and K. Hagino,
  Phys. Rev. C {\bf 76}, 064316  (2007); Phys. Rev. C {\bf 77}, 054309 (2008).
\bibitem{Engel}  J. Engel et al.,  Phys. Rev. C{\bf 60}, 014302 (2000).

\bibitem{M01}M. Matsuo, Nucl. Phys. {\bf A696}, 371 (2001).
\bibitem{Khan02}E. Khan {\it et al.}, Phys. Rev. C{\bf 66}, 024309 (2002).
\bibitem{Paar03}N. Paar {\it et al.}, Phys. Rev. C{\bf 67}, 034312 (2003).
\bibitem{Garrido} E. Garrido et al.,  Phys. Rev. C{\bf 60}, 064312 (1999);
 E. Garrido et al.,  Phys. Rev. C{\bf 63}, 037304 (2001).
\bibitem{Poves98} A. Poves, G. Martinez-Pinedo, Phys. Lett. B430, 203 (1998).

\bibitem{Bertsch09}
G.F. Bertsch and Y. Luo, Phys. Rev. C{\bf 81},
064320 (2010), and references therein. 

\bibitem{Satula97} W. Satula, D. J. Dean, J. Gary, S. Mizutori and W. Nazarewicz,Phys. Lett.{\bf B  B407}, 103 (1997).  \\
W. Satula and R. Wyss, Phys. Lett. {\bf B393}, 1(1997).
  \bibitem{Sakai} T. Wakasa et al., Phys. Rev. C{\bf 55}, 2909 (1997).\\
 M. Ichimura, H.  Sakai and  T. Wakasa, Prog. Part. Nucl. Phys. {\bf 56},
  446 (2006).
 \bibitem{Suzuki}  T. Suzuki, M.  Honma, H\'el\`ene  Mao, T. Otsuka and T. Kajino,
   Phys. Rev. C{\bf 83}, 044619(2011).
\bibitem{Sasano11} M. Sasano et al., Phys. Rev. Lett.{\bf 107}, 202501 (2011).
\bibitem{Moya} E. Moya de Guerra et al., Nucl. Phys. A727, 3 (2003). \\
 S. Fracasso and G. Col\`o, Phys. Rev. C{\bf 76},  044307 (2007). 
\bibitem{Wigner37}  E. P. Wigner, Phys. Rev. {\bf 51}, 106(1937); Phys. Rev. {\bf 56}, 519(1939) .\\
F. Hund, Z. Physik 105, 202(1937). 
\bibitem{Bai09}  C. L. Bai, H. Sagawa, H. Q. Zhang, X. Z. Zhang, G. Col\`o and
 F. R. Xu,  Phys. Rev C{\bf 79}, 041301(R)(2009); Phys. Lett. {\bf B675}, 28(2009).
\bibitem{Bai10}  C. L. Bai, H. Q. Zhang, X. Z. Zhang, F. R. Xu,  H. Sagawa and
   G. Col\`o,   Phys. Rev. Lett. {\bf 105}, 072501 (2010).
\bibitem{Bai11}  C. L. Bai, H. Q. Zhang,  H. Sagawa,  X. Z. Zhang,  G. Col\`o  and  F. R. Xu, Phys. Rev C{\bf 83}, 054316 (2011).
 \bibitem{Lesinski07} T. Lesinski,  M. Bender, K. Bennaceur,
   T. Duguet, and  and J. Meyer, Phys. Rev C{\bf 76}, 014312 (2007).
 \bibitem{SH12}  T. Suzuki and M. Honma, private communications. 
 \bibitem{Niu12} Y. F. Niu, G. Col\`o, M. Brenna. B. F. Bortignon and J. Meng,
  Phys. Rev C{\bf 85}, 034314(2012).  
\bibitem{LW74}
E. Lomon and R. Wilson,
Phys. Rev. C{\bf 9}, 1329 (1974).


 \bibitem{BE} G. F. Bertsch and H. Esbensen,   Ann. Phys.  (NY) {\bf 209}, 327(1991).

\bibitem{EBH97}H. Esbensen, G.F. Bertsch, and K. Hencken,
Phys. Rev. C{\bf 56}, 3054 (1997).

\bibitem{Nicole}  N. Vinh Mau, arXiv:0711.3173v1  (Nov. 2007).
\end{thebibliography}






\end{document}